\documentclass{appolb}
\usepackage{epsfig}
\title{DOUBLE LOGARITHMIC TERMS $ln^2x$ IN THE HEAVY QUARK PRODUCTION
$\sigma(P\bar P \rightarrow h\bar h) - \sigma(PP \rightarrow h\bar h)$
CROSS SECTIONS.}

\author{Dorota Kotlorz $^1$, Andrzej Kotlorz $^2$
\address{$^1$Department of Physics Ozimska 75, $^2$Department of 
Mathematics Luboszycka 3, Technical University of Opole, 
45-370 Opole, Poland, e-mail $^1$: {\tt dstrozik@po.opole.pl}}}

\begin{document}
\pagestyle{plain}
\eqsec
\maketitle

\begin{abstract}
Predictions for the difference of proton-antiproton and proton-proton cross 
sections in the heavy quark production within LO DGLAP analysis together with 
the $ln^2x$ terms resummation are presented. An important role of the double
logarithmic $ln^2x$ corrections in a case of the large rapidity gap between
a quark and an antiquark is discussed.
\end{abstract}


\section{Introduction}

Tevatron, LHC and SSC experiments enable to investigate a new kinematical
region of hadron collisions, where the hadron center-of-mass energy
$\sqrt{s}$ is much larger than the parton momentum transfer $Q$. Large
enough scale $Q^2$ justifies the use of perturbative QCD (PQCD) and
therefore hadron scattering data could be a test not only of the Regge
theory (soft physics) but of different PQCD approaches e.g. DGLAP or BFKL
too. The high energy hadronic processes depending upon the values of the
momentum transfer squared $Q^2$ can be divided on the soft (with small $Q^2$
and small $x$ involved) and the hard ones (at large $Q^2>1 {\rm GeV}^2$). The high
energy behaviour of the total hadronic cross sections in the Regge theory is
described by two contributions: one is close $s^{0.1}$, the other is close
$s^{-0.5}$. The first term $\sim s^{0.1}$ results from soft pomeron exchange
while the second one $\sim s^{-0.5}$ is identified as corresponding to
$\rho, \omega, f_2$ or $a_2$ exchange. Thus the Regge poles theory predicts
a decrease of $\sigma^{tot}(s)$ as $s \rightarrow \infty$. Experimental data
show however that at $s\sim 100 {\rm GeV}^2$ the total hadronic cross sections
have rather a weak energy dependence i.e. they slowly (logarithmically $\sim
ln^2s$) increase with energy at larger energies. On the other side PQCD
predicts very rapid increase of parton distribution functions (PDFs) at
small $x$ (i.e. large $s$) region. In the leading $ln(1/x)$ BFKL approach
one obtains bare Pomeron, which causes that cross sections increase as a
power of energy, what is the evident unitarity violation. Very interesting
both experimentally and theoretically are so called semi-hard processes
(SHP), which involve relatively large transverse momenta $Q^2$ $(Q^2\gg
\Lambda^2 \sim 0.04 {\rm GeV}^2)$ and soft parton collisions with low momentum
fraction $x$ $(x<10^{-2})$ as well. In this way SHP open the description
area in which the Regge theory and PQCD adjoin. Unfortunately, in this
semihard regime there is a uncertainty in determination of the small $x$
dependence of parton distributions. In order to avoid this uncertainty in
predictions of jet production cross sections it is necessary to measure the
two-jet inclusive cross section in hadron collisions with a rapidity gap
between these two jets. In such processes the dominant part is an exchange
of BFKL Pomeron between partons. Events with a rapidity gap i.e. a region of
rapidity in which no particles are found occur if at the parton level the
colour singlet ($q\bar q$ or $gg$) is exchanged in the t-channel. Though the
Pomeron $(gg)$ exchange mechanism is the leading one, the second
contribution $(q\bar q)$ coming from the mesonic reggeons is interesting
too. Analysis of the mesonic reggeon exchange enable to test double
logarithmic effects in PQCD. Elimination of the background origing from
Pomeron, reggeized gluon and odderon in QCD analysis of hadron collisions
gives possibility to determine the appropriate cross sections via valence
quark-antiquark elastic scattering terms. The difference of the cross
sections of $P\bar P$ and $PP$ collisions incorporating $ln^2x$
approximation in valence quark distribution functions is not too small
($\geq 100$ pb) and can be measured e.g. in Tevatron and LHC.

In our paper we show predictions for the heavy quark production cross section
in hadron collisions. Taking into account the difference 
$\sigma(P\bar P \rightarrow h\bar h) - \sigma(PP \rightarrow h\bar h)$ we are
able to investigate the pure mesonic exchange picture modified by PQCD
$ln^2x$ effects. In the next section we shall briefly recall the cross
section formula for the production of heavy quarks in hadron-hadron
collisions. We shall show how to eliminate all contributions except one
origing from the mesonic reggeon exchange. Then in point 3 we shall present
PQCD modification of the Regge model approach in which the $ln^2x$ effects
in valence quark distributions are taken into account. We calculate cross
sections $\sigma(P\bar P \rightarrow h\bar h) - \sigma(PP \rightarrow h\bar h)$
and $d\sigma(P\bar P \rightarrow h\bar h)/d\Delta y
 - d\sigma(PP \rightarrow h\bar h)/d\Delta y$ for heavy quarks charm and
bottom production using GRV input parametrizations of valence quarks and
including the $ln^2x$ effects into the DGLAP evolution of these distribution
functions. All our numerical results we shall present in section 4. We shall
emphasize the role of the rapidity gap in our approach. Finally in
conclusions we shall roughly discuss future experimental hopes for
observation of $ln^2x$ effects in hadron-hadron collisions.

\section{Production of heavy quarks in hadron-hadron collisions. Valence
quark-antiquark elastic scattering as the only contribution to 
$\sigma(P\bar P \rightarrow h\bar h) - \sigma(PP \rightarrow h\bar h)$
cross section.}

The total cross section for heavy quark production in hadron-hadron
collisions via factorization theorem can be written in the form \cite{b1}:
\begin{equation}\label{r2.1}
\sigma (AB \rightarrow h\bar h) =
\sum\limits_{i,j}\int\limits_{\frac{4m_h^2}{s}}^{1} dx_1 p_{i/A}(x_1,Q^2) 
\int\limits_{\frac{4m_h^2}{sx_1}}^{1} dx_2 p_{j/B}(x_2,Q^2)
\int\limits_{t_{min}}^{t_{max}} \frac{d\hat{\sigma}}{d\hat{t}} d\hat{t} 
\end{equation}
where A,B denote colliding hadrons, $h(\bar h)$ is a heavy quark (antiquark)
with mass $m_h$, $i,j$ run over all partons i.e. quarks and gluons: $(i,j) =
(g,g), (q_{\alpha},\bar q_{\alpha}), (\bar q_{\alpha},q_{\alpha}), \alpha =
u,d,s$. $p_{i/A}(x,Q^2)$ is the parton of $i$ sort (quark, antiquark or
gluon) distribution in hadron $A$, $x_{1,2}$ are the longitudinal momenta
fraction of partons with respect to their parent hadrons. $Q^2$ is the
parton momentum transfer squared $(Q^2\approx 4m_h^2)$. $d\hat{\sigma}
/d\hat{t}$ is the differential partonic cross section, which in a case of 
quark-antiquark subprocess in LO approximation has a form \cite{b2}:
\begin{equation}\label{r2.2}
\frac{d\hat{\sigma} (q\bar q \rightarrow h\bar h)}{d\hat{t}} =
\frac{4\pi\alpha_s^2(Q^2)}{9\hat{s}^4}[(\hat{t}-m_h^2)^2 + (\hat{u}-m_h^2)^2 +
2m_h^2\hat{s}]
\end{equation}
$\alpha_s(Q^2)$ is the strong coupling constant; $\hat{s}, \hat{t}, \hat{u}$
are the well known internal Mandelstam kinematic variables \cite{b2} and
\begin{equation}\label{r2.3}
\hat{s} + \hat{t} + \hat{u} = 2m_h^2
\end{equation}
Hence (\ref{r2.2}) takes a form:
\begin{equation}\label{r2.4}
\frac{d\hat{\sigma} (q\bar q \rightarrow h\bar h)}{d\hat{t}} =
\frac{4\pi\alpha_s^2(Q^2)}{9\hat{s}^4}(2m_h^4 + \hat{s}^2 + 2\hat{t}^2 +
2\hat{s}\hat{t} - 4m_h^2\hat{t})
\end{equation}
After integration with respect to $\hat{t}$ one can obtain the total
partonic cross section, which in a case $q\bar{q}$ part has a form:
\begin{equation}\label{r2.5}
\hat{\sigma}^{(q\bar q \rightarrow h\bar h)}(s,Q^2) = 
\int\limits_{t_{min}}^{t_{max}} \frac{d\hat{\sigma}}{d\hat{t}} d\hat{t}
\end{equation}
The lower and upper limits in the above integral are:
\begin{equation}\label{r2.6}
t_{min} = \frac{-\hat{s}}{4}(1+\beta)^2 
\end{equation}
\begin{equation}\label{r2.7}
t_{min} = \frac{-\hat{s}}{4}(1-\beta)^2 
\end{equation}
with
\begin{equation}\label{r2.8}
\beta^2 = 1 - \frac{4m_h^2}{\hat{s}} 
\end{equation}
and hence one can simply obtain:
\begin{equation}\label{r2.9}
\hat{\sigma}^{(q\bar q \rightarrow h\bar h)}(s,Q^2) = \frac{4\pi
\alpha_s^2(Q^2)}{27\hat{s}}\beta (3-\beta^2)
\end{equation}
The total cross section for heavy quark production is of course dominated by
the gluon-gluon $gg \rightarrow h\bar{h}$ subprocess and if we want to deal
only with a part coming from the quark-antiquark scattering we should
consider suitable difference of cross sections. As it has been exactly
described in \cite{b3}, the simplest way to eliminate the strong background
due to the exchange of Pomeron, reggeized gluon and oderon is to regard
$P\bar{P}$ (proton-antiproton) and $PP$ (proton-proton) collisions. All these
mentioned above backgrounds cancel if one takes the difference of the cross
sections $\sigma (P\bar{P}) - \sigma (PP)$. Then the difference of the 
$P\bar{P}$ and $PP$ cross sections for the heavy quark production is purely
represented by valence quark-antiquark elastic scattering (with mesonic
reggeon exchange in the Regge theory). One can write the differential cross 
section as \cite{b3}:
\begin{eqnarray}\label{r2.10}
\frac{d\sigma (P\bar{P} \rightarrow h\bar{h})}{dx_1 dx_2 d\hat{t}} - 
\frac{d\sigma (PP \rightarrow h\bar{h})}{dx_1 dx_2 d\hat{t}}
= \frac{1}{2}\sum\limits_{f} [q_{f/P}(x_1,Q^2) \Delta\bar{q}_{f/P}(x_2,Q^2)
\nonumber \\
+ \bar{q}_{f/P}(x_1,Q^2) \Delta q_{f/P}(x_2,Q^2)]\frac{d\hat{\sigma}}{d\hat{t}}
\nonumber\\
\end{eqnarray}
with
\begin{equation}\label{r2.11}
\Delta q_{f/P}(x,Q^2) = q_{f/\bar{P}}(x,Q^2) - q_{f/P}(x,Q^2)
\end{equation}
\begin{equation}\label{r2.12}
\Delta\bar{q}_{f/P}(x,Q^2) = \bar{q}_{f/\bar{P}}(x,Q^2) - \bar{q}_{f/P}(x,Q^2)
\end{equation}
By summation over flavours $f$ in parent hadrons ($P$ or $\bar{P}$) in
(\ref{r2.10}) one gets:
\begin{equation}\label{r2.13}
\frac{d\sigma (P\bar{P} \rightarrow h\bar{h})}{dx_1 dx_2 d\hat{t}} - 
\frac{d\sigma (PP \rightarrow h\bar{h})}{dx_1 dx_2 d\hat{t}} = 
\frac{1}{2}\sum\limits_{f=val} q_{f/P}(x_1,Q^2) \bar{q}_{f/\bar{P}}(x_2,Q^2)
\frac{d\hat{\sigma}}{d\hat{t}} 
\end{equation}
Because
\begin{equation}\label{r2.14}
q_{val/P}(x,Q^2) =  \bar{q}_{val/\bar{P}}(x,Q^2)
\end{equation}
one has:
\begin{eqnarray}\label{r2.15}
\frac{d\sigma (P\bar{P} \rightarrow h\bar{h})}{dx_1 dx_2 d\hat{t}} - 
\frac{d\sigma (PP \rightarrow h\bar{h})}{dx_1 dx_2 d\hat{t}} = 
\frac{1}{2} [u_{val}(x_1,Q^2) u_{val}(x_2,Q^2) \nonumber\\
+ d_{val}(x_1,Q^2) d_{val}(x_2,Q^2)]\frac{d\hat{\sigma}}{d\hat{t}}
\end{eqnarray}
or after integration with respect to $\hat{t}$:
\begin{eqnarray}\label{r2.16}
\frac{d\sigma (P\bar{P} \rightarrow h\bar{h})}{dx_1 dx_2} - 
\frac{d\sigma (PP \rightarrow h\bar{h})}{dx_1 dx_2} = 
\frac{2\pi \alpha_s^2(Q^2)}{27\hat{s}}\beta (3-\beta^2) \nonumber \\
\times [u_{val}(x_1,Q^2) u_{val}(x_2,Q^2) + d_{val}(x_1,Q^2) d_{val}(x_2,Q^2)]
\end{eqnarray}
where $\beta^2$ is defined by (\ref{r2.8})
and
\begin{equation}\label{r2.17}
Q^2\approx 4m_h^2
\end{equation}
Then the difference of the total cross sections is:
\begin{eqnarray}\label{r2.18}
[\sigma (P\bar{P} \rightarrow h\bar{h}) - \sigma (PP \rightarrow
h\bar{h})](s)
 = \frac{2\pi \alpha_s^2(Q^2)}{27}
\int\limits_{\frac{4m_h^2}{s}}^{1} dx_1
\int\limits_{\frac{4m_h^2}{sx_1}}^{1} dx_2\frac{\beta (3-\beta^2)}{\hat{s}}
\nonumber \\
\times [u_{val}(x_1,Q^2) u_{val}(x_2,Q^2)
 + d_{val}(x_1,Q^2) d_{val}(x_2,Q^2)] \nonumber \\
\end{eqnarray}
with internal kinematical variable $\hat{s}$:
\begin{equation}\label{r2.19}
\hat{s} = sx_1x_2
\end{equation}
In the formula (\ref{r2.18}) we have contributions from all possible
rapidity gaps $\Delta y$ between a quark and an antiquark:
\begin{equation}\label{r2.20}
\Delta y = y(h)-y(\bar{h})\approx ln\frac{\hat{s}}{4m_h^2}
\end{equation}
and hence
\begin{equation}\label{r2.21}
0\leq\Delta y\leq ln\frac{s}{4m_h^2}
\end{equation}
In a case of Tevatron with $\sqrt{s}=1.8 {\rm TeV}$ and for $c\bar{c}$ production
with $m_c=1.43 {\rm GeV}$ theoretically
\begin{equation}\label{r2.22}
0\leq\Delta y\leq 13
\end{equation}
However the interesting double logarithmic $ln^2x$ effects are important
when the rapidity gap $\Delta y$ is large:
\begin{equation}\label{r2.23}
(\Delta y)^2 = ln^2\frac{\hat{s}}{4m_h^2}\sim ln^2x
\end{equation}
To see this one should take into consideration the differential cross
section:
\begin{eqnarray}\label{r2.24}
\frac{d\sigma (P\bar{P} \rightarrow h\bar{h})}{d\Delta y} - 
\frac{d\sigma (PP \rightarrow h\bar{h})}{d\Delta y} = 
\frac{2\pi \alpha_s^2(Q^2)}{27s}\int\limits_{\frac{4m_h^2e^{\Delta
y}}{s}}^{1}\frac{dx_1}{x_1}\beta (3-\beta^2) \nonumber\\
\times[u_{val}(x_1,Q^2) u_{val}(\frac{4m_h^2e^{\Delta y}}{s},Q^2) + d_{val}(x_1,Q^2)
d_{val}(\frac{4m_h^2e^{\Delta y}}{s},Q^2)] \nonumber \\
\end{eqnarray}
with $\beta^2$ as a function of $\Delta y$:
\begin{equation}\label{r2.25}
\beta^2 = 1 - e^{-\Delta y}
\end{equation}
From (\ref{r2.25}) one can simply notice that for a large $\Delta y$
$\beta^2$ is close 1.0:
\begin{equation}\label{r2.26}
\beta^2 = 1~~ as~~\Delta y\longrightarrow \infty
\end{equation}
and hence (\ref{r2.24}) for a large $\Delta y$ can be expressed as:
\begin{eqnarray}\label{r2.27}
\frac{d\sigma (P\bar{P} \rightarrow h\bar{h})}{d\Delta y} - 
\frac{d\sigma (PP \rightarrow h\bar{h})}{d\Delta y} = 
\frac{4\pi \alpha_s^2(Q^2)}{27s}\int\limits_{\frac{4m_h^2e^{\Delta y}}{s}}^{1}
\frac{dx_1}{x_1} \nonumber \\
\times [u_{val}(x_1,Q^2) u_{val}(\frac{4m_h^2e^{\Delta y}}{s},Q^2) + 
d_{val}(x_1,Q^2) d_{val}(\frac{4m_h^2e^{\Delta y}}{s},Q^2)]
\end{eqnarray} 
According to (\ref{r2.23}), with larger rapidity gap $\Delta y$ the $ln^2x$
effects in the $\sigma(P\bar P \rightarrow h\bar h) - \sigma(PP \rightarrow 
h\bar h)$ would be better visible. These effects we introduce via suitable
modification in DGLAP LO evolution equations for valence quarks $u_{val}$ and
$d_{val}$. The perturbative QCD approach with an appearance of the double
logarithmic terms $ln^2x$ can modify the Regge model. We discuss this in the
next section.

\section{Double logarithmic effects $ln^2x$ for flavour nonsiglet parton
distribution functions.}

The high energy $s$ limit corresponds by definition to the Regge limit. In
the Regge pole model \cite{b4} an amplitude of high energy elastic scattering
$T(s,t)\sim s^{\alpha (t)}$, where $\alpha (t)$ is the Regge trajectory,
which is equal to spin $J$ of the corresponding particle at $t=M^2$ ($M$ is
the mass of the exchanged particle). Hence the total interaction cross
section, which by the optical theorem is connected to $ImT(s,0)$ can be
written as a sum of the Regge poles contributions:
\begin{equation}\label{r3.1}
\sigma^{tot}(s) = \sum\limits_{k} b_k(0) s^{\alpha_k (0)-1}
\end{equation}
where $\alpha_k(0)$ are the intercepts of the Regge trajectories and $b_k$
denote the couplings. The high energy behaviour of the total hadronic cross
section can be described by two parts: first contribution is so called soft
pomeron with intercept $\sim 1.08$ and the second one are the leading meson
Regge trajectories with intercept $\alpha_R(0)\approx 0.5$. These reggeons
correspond to $\rho, \omega, f$ or $a_2$ mesons exchanges. The flavour
nonsinglet part e.g. valence quark distribution functions or
$F_2^{NS}=F_2^P-F_2^N$ is governed at small $x$ (i.e. at high energy
$s=Q^2(1/x-1)$) by $a_2$ reggeon:
\begin{equation}\label{r3.2}
Regge:~~~q_{val}(x)\sim x^{-\alpha_{a_2}(0)}
\end{equation}
with $\alpha_{a_2}(0)\approx 0.5$. This behaviour is stable against the
leading order DGLAP QCD evolution \cite{b5}. In other words the PQCD $Q^2$
evolution behaviour of $q_{val}$ at high energy is screened by the leading
Regge contribution (\ref{r3.2}). But there is a novel effect concerning the
flavour nonsinglet functions which is an appearance of the double
logarithmic terms $ln^2x$ \cite{b6}, \cite{b7}. For not too small values of 
the QCD coupling $\alpha_s$ this contribution is approximately the same 
(or even greater) in comparison to the contribution of the $a_2$ Regge pole:
\begin{equation}\label{r3.3}
ln^2x:~~~q_{val}(x)\sim x^{-\bar\omega}
\end{equation}
and
\begin{equation}\label{r3.4}
\bar\omega = 2\sqrt{\bar\alpha_s}
\end{equation}
where
\begin{equation}\label{r3.5}
\bar\alpha_s = \frac{2\alpha_s}{3\pi}
\end{equation}
At a typical value of $\alpha_s(Q^2=4m_h^2)$ for heavy quark production:
\begin{equation}\label{r3.6}
\alpha_s(Q^2=4m_c^2) = 0.30
\end{equation}
\begin{equation}\label{r3.7}
\alpha_s(Q^2=4m_b^2) = 0.21
\end{equation}
$\bar\omega$ becomes very close to $\alpha_{a_2}$:
\begin{equation}\label{r3.8}
\bar\omega(\alpha_s(Q^2=4m_c^2) \approx 0.5
\end{equation}
\begin{equation}\label{r3.9}
\bar\omega(\alpha_s(Q^2=4m_b^2) \approx 0.4
\end{equation}
It must be however emphasized that even in the case of the PQCD singularity
generated by the double logarithmic $ln^2x$ resummation, this $ln^2x$
behaviour of flavour nonsinglet parton distributions is usually hidden
behind leading $a_2$ reggeon exchange contribution. Only in the case of
large rapidity gap $\Delta y$, where the intercept of the Regge trajectory
$\alpha_{a_2}(0)$ becomes close to 0 or even negative, it is possible to
separate from the Regge background the $ln^2x$ behaviour of the considered
cross sections. In our analysis we calculate the difference of cross
sections (\ref{r2.18}) and (\ref{r2.24}) using $u_{val}(x,Q^2)$ and 
$d_{val}(x,Q^2)$ obtained from DGLAP LO evolution equations, incorporating
the double logarithmic effects $ln^2x$ as well. This method combining DGLAP
LO and $ln^2x$ approaches in the case of polarized flavour nonsinglet
functions was presented in \cite{b8}, \cite{b9}, \cite{b10}. For unpolarized 
ones, which we now investigate, the suitable equations are the same, only the 
input parametrizations $q_{val}(x,k_0^2)$ are different. In our approach we 
expect that the cross sections (\ref{r2.18}) and particularly this for the 
large rapidity gap (\ref{r2.24}) are governed by $ln^2x$ terms i.e. by the 
valence quark ladder, shown in Fig.1.
%***Fig.1***
\begin{figure}[ht]
\begin{center}
\includegraphics[width=70mm]{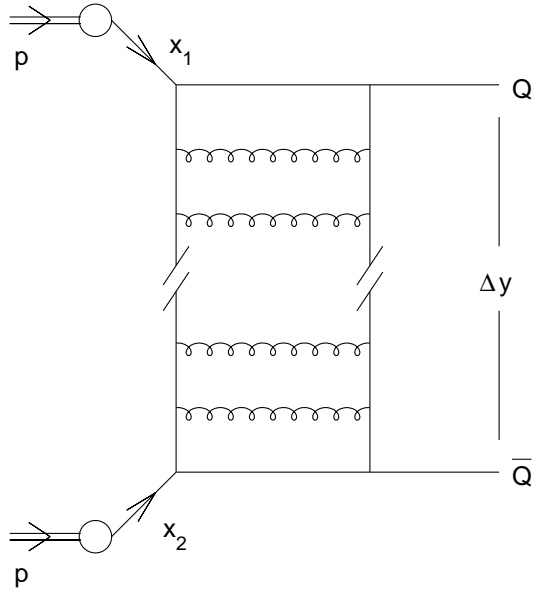}
\caption{A ladder diagram generating double logarithmic $\ln^2(1/x)$
terms in the flavour nonsinglet part of heavy quark production cross
section with large rapidity gap $\Delta y$.}
\end{center}
\end{figure}
The equation taking into account both DGLAP LO evolution and $ln^2x$ effects
for $q_{val}(x,Q^2)$ functions has a form \cite{b9}:
\begin{eqnarray}\label{r3.10}
f(x,k^2)&=&f^{(0)}(x,k^2)+\frac{2\alpha_s(k^2)}{3\pi}
\int\limits_x^1\frac{dz}{z}\int\limits_{k_0^2}^{k^2/z}
\frac{dk'^2}{k'^2}f(\frac{x}{z},k'^2)\nonumber\\
&+&\frac{\alpha_s(k^2)}{2\pi}
\int\limits_{k_0^2}^{k^2}\frac{dk'^2}{k'^2}[\frac43
\int\limits_x^1\frac{dz}{z}\frac{(z+z^2)f(x/z,k'^2)-2zf(x,k'^2)}{1-z}
\nonumber\\
&+&(\frac12+\frac83\ln (1-x))f(x,k'^2)]
\end{eqnarray}
where
\begin{eqnarray}\label{r3.11}
f^{(0)}(x,k^2)&=&\frac{\alpha_s(k^2)}{2\pi}
[\frac43\int\limits_x^1\frac{dz}{z}\frac{(1+z^2)q_{val}^{(0)}(x/z)
-2zq_{val}^{(0)}(x)}{1-z}\nonumber\\
&+&(\frac12+\frac83\ln (1-x))q_{val}^{(0)}(x)]
\end{eqnarray}
The unintegrated distribution $f$ in the equation (\ref{r3.10})
are related to the $q_{val}(x,Q^2)$ via
\begin{equation}\label{r3.12}
q_{val}(x,Q^2)=q_{val}^{(0)}(x)+\int\limits_{k_0^2}^{Q^2(1/x-1)}
\frac{dk^2}{k^2}f\left(x(1+\frac{k^2}{Q^2}),k^2\right)
\end{equation}
where
\begin{equation}\label{r3.13}
q_{val}^{(0)}(x)\equiv q_{val}(x,k_0^2)=
\int\limits_0^{k_0^2}\frac{dk^2}{k^2}f(x,k^2)
\end{equation}
In above equations $q_{val}(x,Q^2)$ denotes $u_{val}(x,Q^2)$ and
$d_{val}(x,Q^2)$ distributions as well. In our calculations we use LO fitted
GRV \cite{b11} input parametrizations $q_{val}(x,k_0^2)$:
\begin{equation}\label{r3.14}
k_0^2 = 1 {\rm GeV}^2,~~~\Lambda (n_f=4) = 232 {\rm MeV}
\end{equation}
\begin{eqnarray}\label{r3.15}
GRV&:&u_{val}(x,k_0^2=1~{\rm GeV}^2)=2.872x^{-0.427}\nonumber\\
&\times&(1-0.583x^{0.175}+1.723x+3.435x^{3/2})(1-x)^{3.486}
\end{eqnarray}
\begin{eqnarray}\label{r3.16}
GRV&:&d_{val}(x,k_0^2=1~{\rm GeV}^2)=0.448x^{-0.624}\nonumber\\
&\times&(1+1.195x^{0.529}+6.164x+2.726x^{3/2})(1-x)^{4.215}
\end{eqnarray}
The numerical results for (\ref{r2.18}) and (\ref{r2.24}) cross sections we
present in the next point.

\section{Numerical predictions for $\sigma (P\bar{P} \rightarrow h\bar{h}) 
- \sigma (PP \rightarrow h\bar{h})(s)$ and $\frac{d\sigma (P\bar{P} 
\rightarrow h\bar{h})}{d\Delta y} - \frac{d\sigma (PP \rightarrow h\bar{h})}
{d\Delta y}$ incorporating DGLAP LO evolution and the $ln^2x$ effects
in $q_{val}$ functions.}

Our numerical results based on GRV (\ref{r3.14})-(\ref{r3.16}) input
parametrizations are presented in Figs.2-7 and in Table I. In Fig.2 we plot
$xq_{val}$ input parametrizations for $u$ and $d$ valence quarks in proton at 
$k_0^2=1 {\rm GeV}^2$ together with their LO DGLAP + $ln^2x$ predictions at
$Q^2=10 {\rm GeV}^2$. Figs.3,4 show the difference of total cross sections 
(\ref{r2.18}) which incorporates $ln^2x$ terms in $q_{val}(x,Q^2)$
functions. Our calculations concern the charm quark (Fig.3) and the bottom
quark (Fig.4) production. In Fig.5 we compare LO DGLAP + $ln^2x$ prediction
for the cross section (\ref{r2.24}) with pure LO DGLAP one at large $\Delta y
=4.0$. In Figs.6,7 we plot the difference of differential cross sections 
(\ref{r2.24}) with LO DGLAP + $ln^2x$ $q_{val}(x,Q^2)$ for both charm and bottom
production. The plots in Figs.6,7 are performed for different large rapidity gap
$\Delta y \geq 2.0$. All plots in Figs.3-7 are presented as a function of the 
center mass energy $\sqrt{s}$. Finally in Table I we present the differential
cross section 
\begin{eqnarray}\label{r4.1}
x_1 x_2 [\frac{d\sigma (P\bar{P} \rightarrow h\bar{h})}{dx_1 dx_2} - 
\frac{d\sigma (PP \rightarrow h\bar{h})}{dx_1 dx_2}] = 
\frac{4\pi \alpha_s^2(Q^2)}{27 s}\nonumber\\
\times [u_{val}(x_1,Q^2) u_{val}(x_2,Q^2) + d_{val}(x_1,Q^2) d_{val}(x_2,Q^2)]
\end{eqnarray}
for large rapidity gaps at different $\sqrt{s}$ and suitable $x_1=x_2$:
\begin{equation}\label{r4.2}
x_1 = x_2 = \frac{2m_h}{\sqrt{s}} e^{\frac{\Delta y}{2}}
\end{equation}
%***Fig.2***
\begin{figure}[ht]
\begin{center}
\includegraphics[width=105mm]{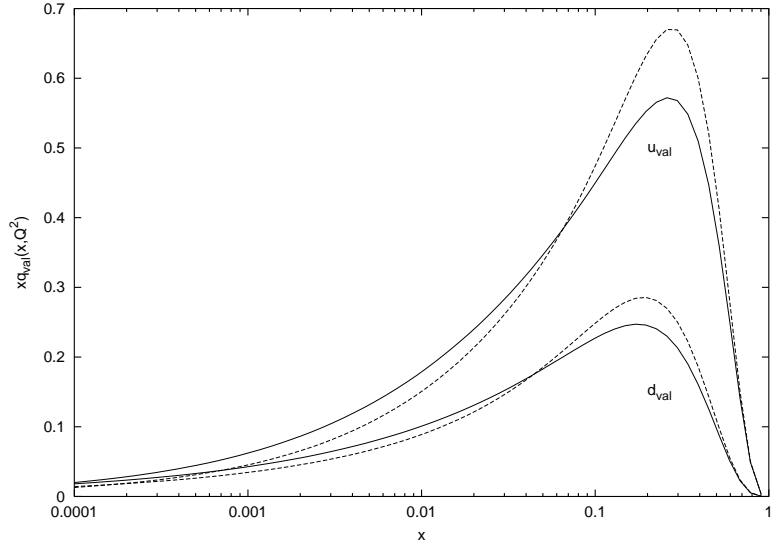}
\caption{Valence quark distribution functions $xu_{val}$ and $xd_{val}$ at
input scale $k_0^2 = 1 {\rm GeV}^2$ - dashed and at $Q^2 = 10 {\rm GeV}^2$ -
solid.}
\end{center}
\end{figure}
%***Fig.3***
\begin{figure}[ht]
\begin{center}
\includegraphics[width=105mm]{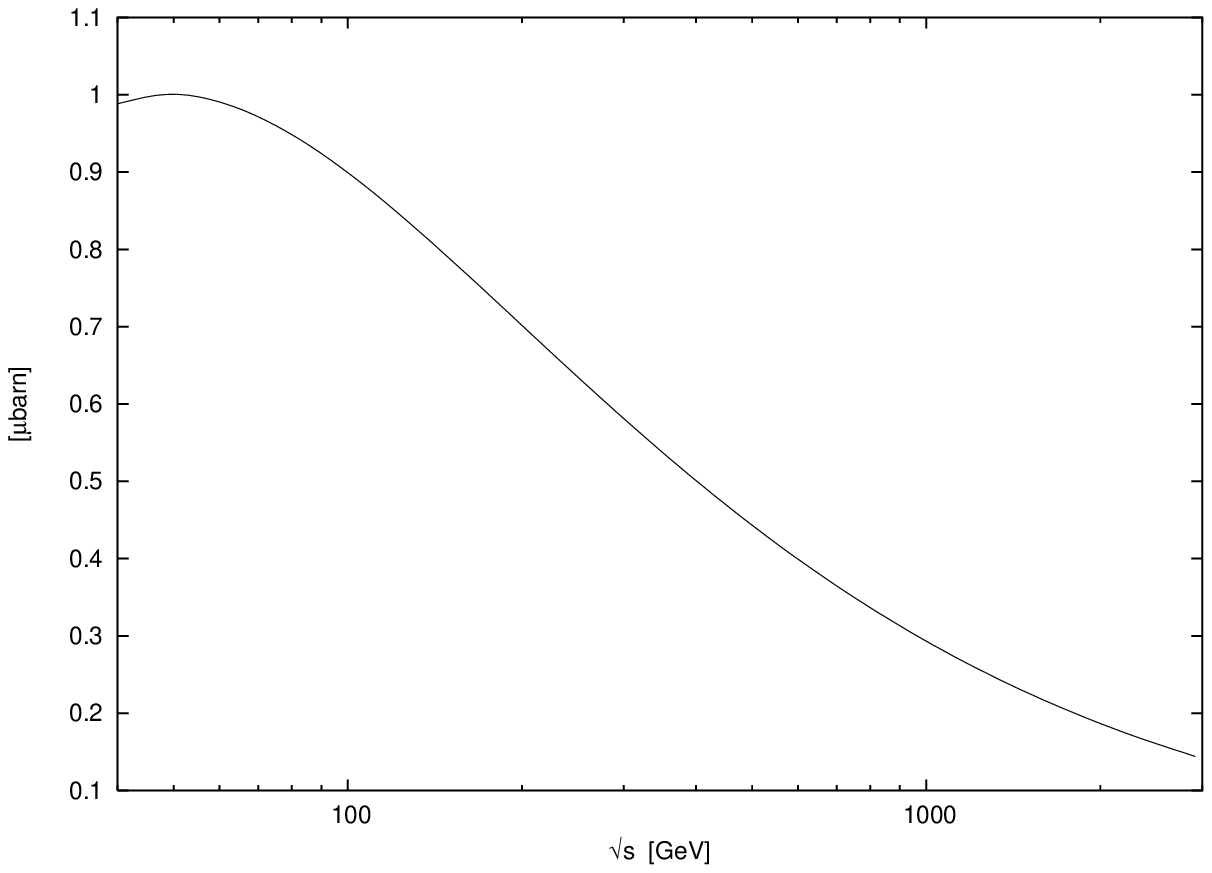}
\caption{Predictions for $\sigma (P\bar{P} \rightarrow h\bar{h}) 
- \sigma (PP \rightarrow h\bar{h})(s)$ incorporating LO DGLAP + $ln^2x$
$q_{val}(x,Q^2)$ for the charm quark production.}
\end{center}
\end{figure}
%***Fig.4***
\begin{figure}[ht]
\begin{center}
\includegraphics[width=100mm]{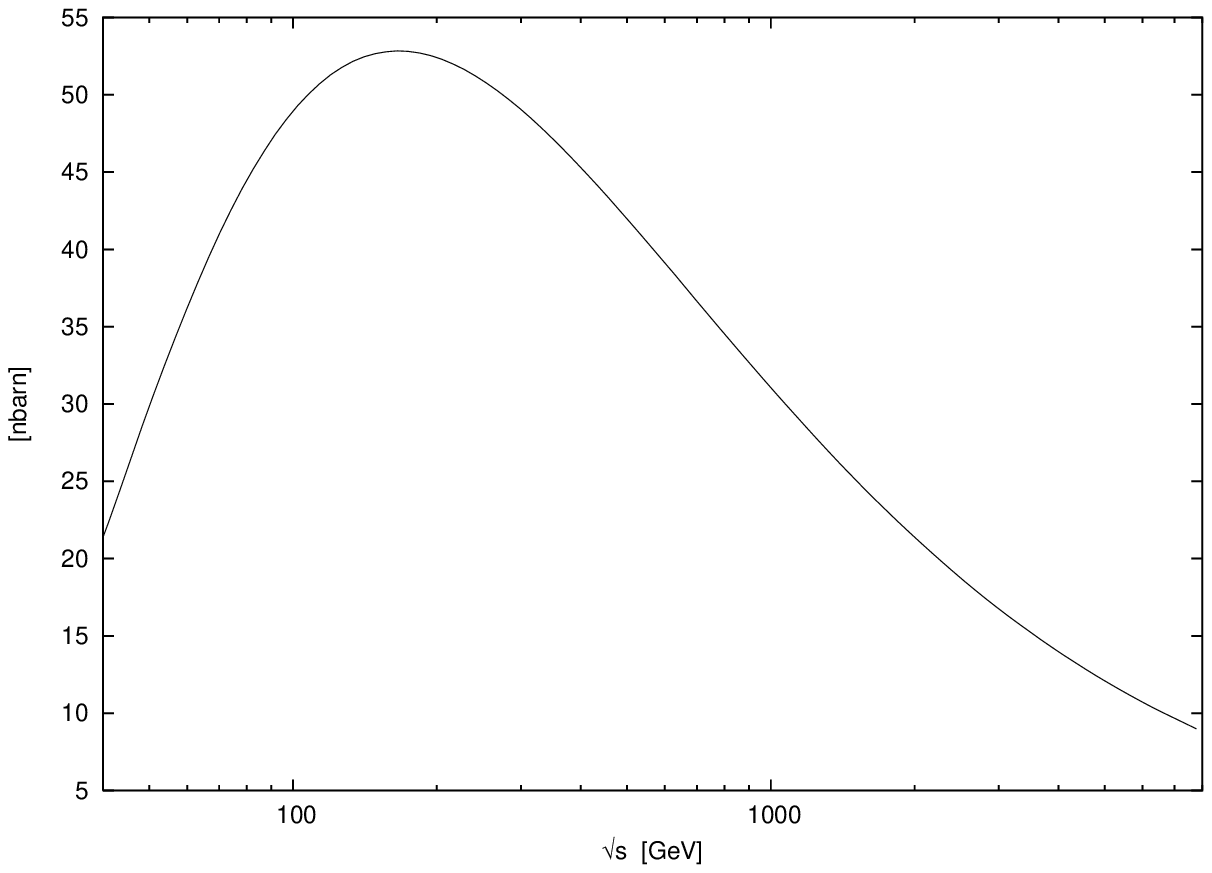}
\caption{Predictions for $\sigma (P\bar{P} \rightarrow h\bar{h}) 
- \sigma (PP \rightarrow h\bar{h})(s)$ incorporating LO DGLAP + $ln^2x$
$q_{val}(x,Q^2)$ for the bottom quark production.}
\end{center}
\end{figure}
%***Fig.5***
\begin{figure}[ht]
\begin{center}
\includegraphics[width=100mm]{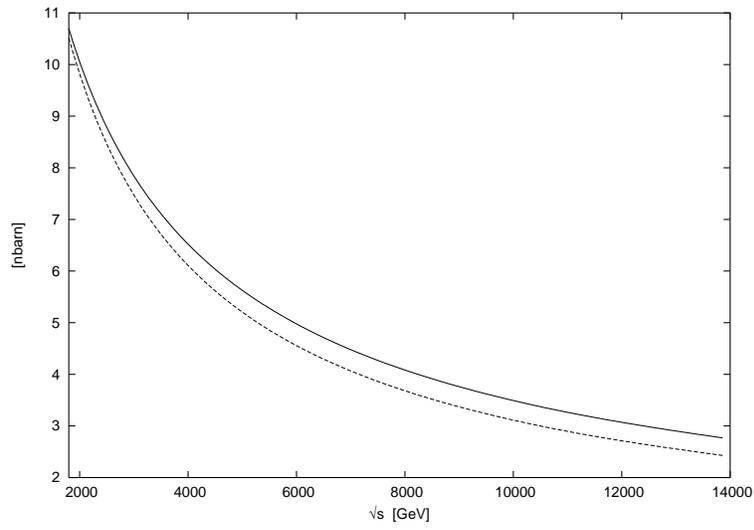}
\caption{Comparison of pure LO DGLAP predictions for $\frac{d}{d\Delta{y}}
[\sigma (P\bar{P} \rightarrow h\bar{h}) - \sigma (PP \rightarrow h\bar{h})]$
- dashed with LO DGLAP + $ln^2x$ ones - solid at $\Delta{y} = 4.0$ for the 
charm quark production.}
\end{center}
\end{figure}
%***Fig.6***
\begin{figure}[ht]
\begin{center}
\includegraphics[width=95mm]{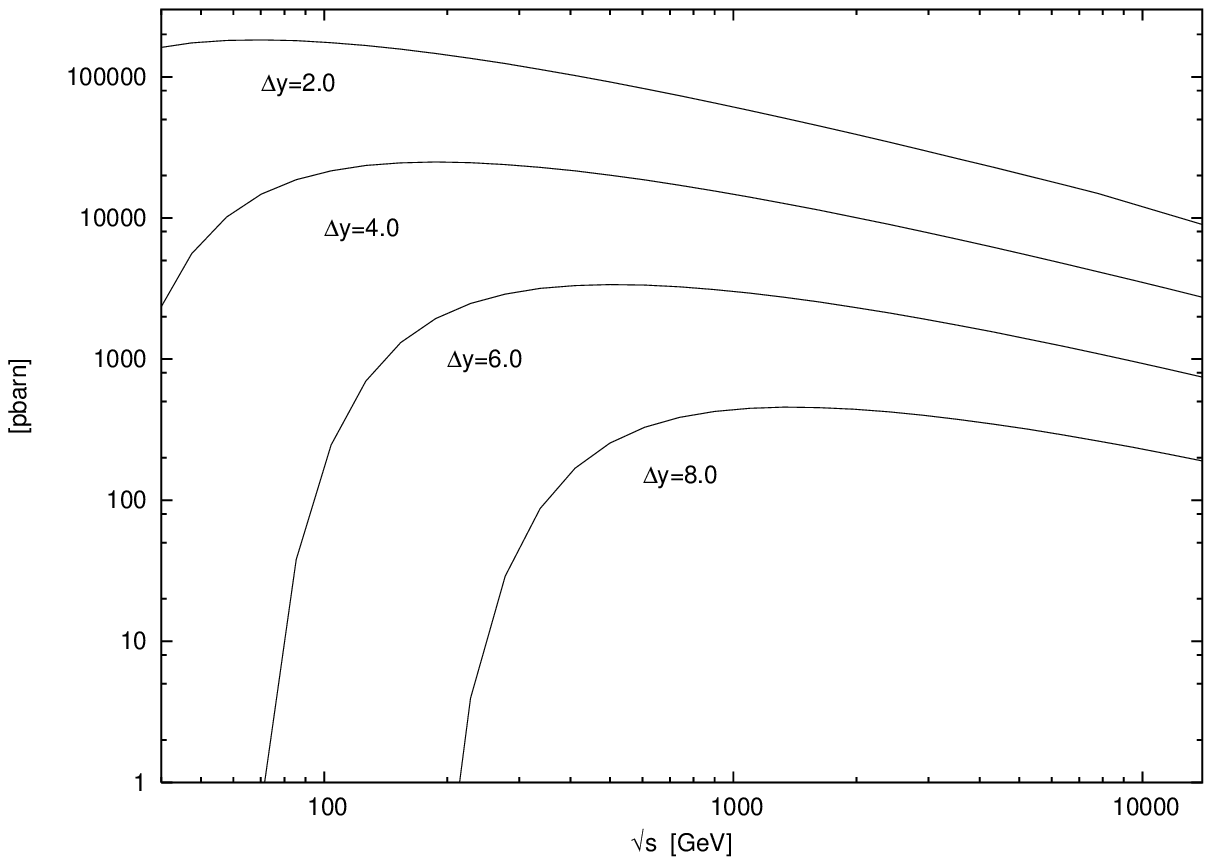}
\caption{Differential cross section $\frac{d}{d\Delta{y}}
[\sigma (P\bar{P} \rightarrow h\bar{h}) - \sigma (PP \rightarrow h\bar{h})]$
incorporating LO DGLAP + $ln^2x$ $q_{val}(x,Q^2)$ for the charm quark 
production at different large $\Delta{y}$.}
\end{center}
\end{figure}
%***Fig.7***
\begin{figure}[ht]
\begin{center}
\includegraphics[width=95mm]{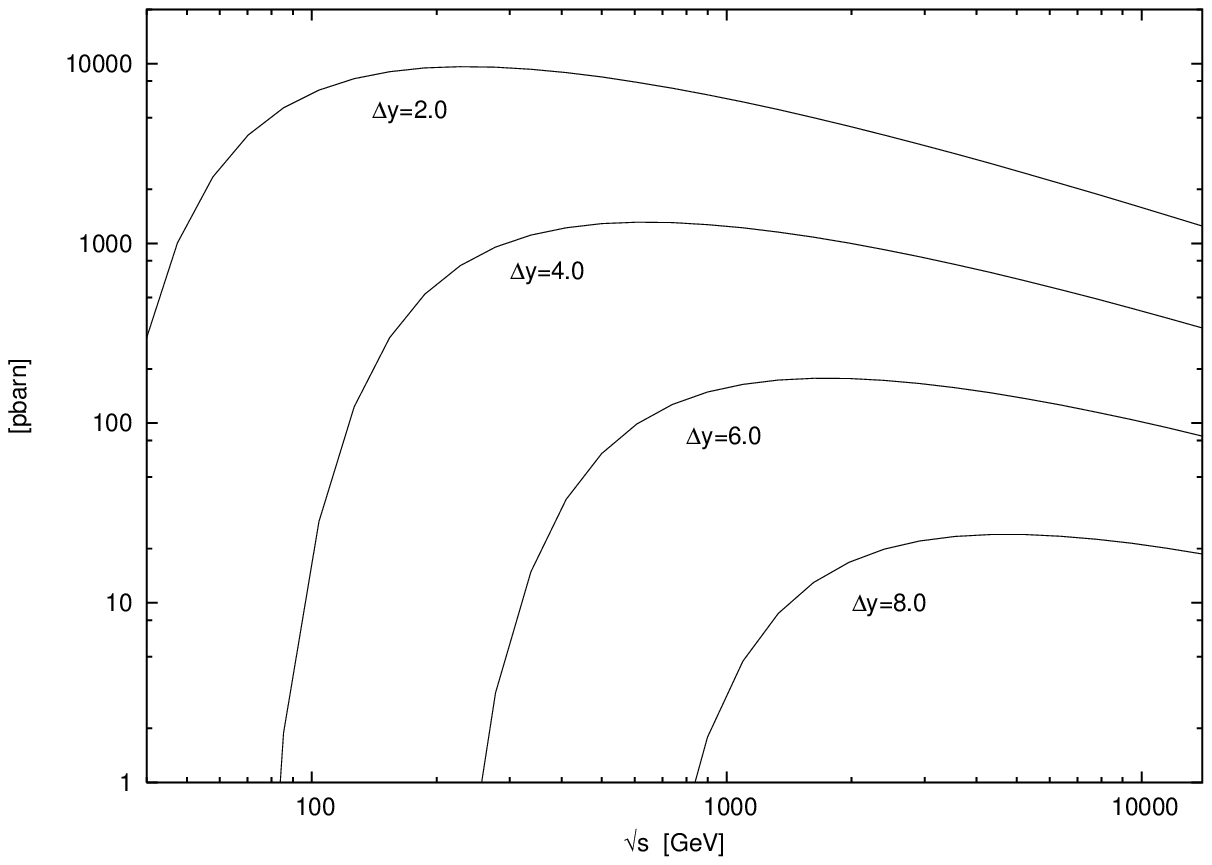}
\caption{Differential cross section $\frac{d}{d\Delta{y}}
[\sigma (P\bar{P} \rightarrow h\bar{h}) - \sigma (PP \rightarrow h\bar{h})]$
incorporating LO DGLAP + $ln^2x$ $q_{val}(x,Q^2)$ for the bottom quark 
production at different large $\Delta{y}$.}
\end{center}
\end{figure}
%***Table I***
\begin{table}[ht]
\begin{center}
\begin{tabular}{|c|c|c|c|}
\hline\hline
$x_1=x_2$ &  $\sqrt(s) [{\rm GeV}]$ & $\Delta{y}$ & ${\rm F}$ [pb] \\ \hline
0.54  &  39     & 4.0 & 3776 \\
0.034 & 630     & 4.0 & 3811 \\
0.091 & 630     & 6.0 & 1090 \\
0.25  & 630     & 8.0 & 236 \\
0.012 & 1800    & 4.0 & 1628 \\
0.032 & 1800    & 6.0 & 496 \\
0.087 & 1800    & 8.0 & 143 \\
0.0015 & 14000  &  4.0 & 277 \\
0.0041 & 14000  &  6.0 & 90 \\
0.011  & 14000  &  8.0 & 29 \\
\hline
\end{tabular}
\caption{\label{tab:par}differential cross section $ {\rm F}\equiv 
x_1 x_2 [\frac{d\sigma (P\bar{P} \rightarrow h\bar{h})}{dx_1 dx_2} - 
\frac{d\sigma (PP \rightarrow h\bar{h})}{dx_1 dx_2}]$ for a charm quark at
large $\Delta{y}\geq 4.0$ and different $\sqrt{s}$.}
\end{center}
\end{table}

From Figs.3-4 one can read the shape of the cross sections difference 
$\sigma (P\bar{P} \rightarrow h\bar{h}) - \sigma (PP \rightarrow h\bar{h})(s)$
as a function of $\sqrt{s}$. An initial increase of its value with the
increasing value of $\sqrt{s}$ results from the energy threshold for heavy
quark production:
\begin{equation}\label{r4.3}
s\geq \hat{s} \geq 4m_h^2
\end{equation}
where $\hat{s}$ is defined in (\ref{r2.19}). When the energy becomes larger
($\sqrt{s}> 50 {\rm GeV}$ for charm and $\sqrt{s}> 150 {\rm GeV}$ for bottom
production) because of a suppression factor $s^{-1}$ in the cross section
formula (\ref{r2.18}), $\sigma (P\bar{P}) - \sigma (PP)$ decreases with the
further increasing of energy $\sqrt{s}$. More interesting is analysis of the
differential cross section $\frac{d\sigma (P\bar{P} 
\rightarrow h\bar{h})}{d\Delta y} - \frac{d\sigma (PP \rightarrow h\bar{h})}
{d\Delta y}$. When the rapidity gap $\Delta{y}$ between a quark and an antiquark
defined in (\ref{r2.20}) is large, according to (\ref{r2.23}) high order
corrections $\sim (\alpha_s ln^2x)^n$ are important. It is shown in Fig.5
where we compare the LO DGLAP + $ln^2x$ predictions for the cross section 
(\ref{r2.24}) with pure LO DGLAP ones as a function of $\sqrt{s}$ at large 
rapidity gap $\Delta y=4.0$. In the high energy region $\sqrt{s}\in(1800 {\rm GeV};
14 {\rm TeV})$ LO DGLAP + $ln^2x$ $q_{val}$ give larger value of the cross section 
$\frac{d}{d\Delta{y}}[\sigma (P\bar{P}) - \sigma (PP)]$ than the pure LO
DGLAP $q_{val}$ approach. This results simply from larger values of
$q_{val}$ distribution functions in LO DGLAP + $ln^2x$ approach at small $x$
($x\leq 10^{-2}$) in comparison to those obtained via LO DGLAP (or even
NLO) evolution method. However double logarithmic terms $ln^2x$ play a
significant role not only in the very small $x$ region. Indeed, for large
$\Delta{y}$ one deals with not too small $x$ in the integrals (\ref{r2.24})
or (\ref{r2.27}) because the lower limit for $x$ is
\begin{equation}\label{r4.4}
x_{low} =  \frac{4m_h^2}{s} e^{\Delta{y}}
\end{equation}
In a case of a very large rapidity gap e.g. $\Delta{y} = 8.0$ at Tevatron
energy $\sqrt{s} = 1800 {\rm GeV}$ and for $m_h = m_c$ one should know
parton distributions at $x\geq 10^{-2}$ instead of $x\geq 3\cdot
10^{-6}$ ($\Delta{y} = 0$). It could be very convenient because the very
small $x$ region is still experimentally unattainable. Large rapidity gap 
$\Delta{y}$ implies however rapid decrease of suitable cross sections. So 
one should minimize $\hat{s}$ which is a suppression factor ($\hat{s}^{-1}$) 
keeping still large enough $\Delta{y}\sim ln\hat{s}$. Our plots in Figs.6-7 
exhibit that for larger $\Delta{y}$ the differential cross section 
$\frac{d}{d\Delta{y}}[\sigma (P\bar{P}) - \sigma (PP)]$ can change by almost
2 (3) orders of magnitude from 81 nb (8 nb) at $\sqrt{s} = 630 {\rm GeV}$ and 
$\Delta{y} = 2.0$ to 1.2 nb (10 pb) at $\Delta{y} = 7.0$ and the same energy 
in a charm (bottom) production. For one value of $\Delta{y} = 4.0$ we can find 
$\frac{d}{d\Delta{y}}[\sigma (P\bar{P}) - \sigma (PP)]$ for a charm (bottom)
case in the range from 22 nb (28 pb) at $\sqrt{s}=100 {\rm GeV}$ to 2.8 nb 
(340 pb) at $\sqrt{s}=14 {\rm TeV}$. Table I shows values of double 
differential cross section 
$x_1 x_2 [\frac{d\sigma (P\bar{P} \rightarrow h\bar{h})}{dx_1 dx_2} - 
\frac{d\sigma (PP \rightarrow h\bar{h})}{dx_1 dx_2}]$ for a charm quark at
large $\Delta{y}\geq 4.0$ and different $\sqrt{s}$. These DGLAP + $ln^2x$
predictions which are equal from 29 pb ($\Delta{y} = 8.0$, $\sqrt{s} = 14
{\rm TeV}$) to 3.8 nb ($\Delta{y} = 4.0$, $\sqrt{s} = 39 {\rm GeV}$) as not too
small should probably be measurable.

\section{Summary.}

Difference of two cross sections $\sigma (P\bar{P}) - \sigma (PP)]$ for
heavy quark production is in the Regge theory represented by pure mesonic
reggeon exchange process. In this way only valence quarks and antiquarks are
taken into account in this hadron-hadron collisions. Distribution functions
of valence quarks $q_{val}(x,Q^2)$ which via factorization theorem enter to
a suitable formula for quark-antiquark elastic scattering in cross sections
can be obtained within PQCD approach. In our paper we have used a
perturbative method which incorporates LO DGLAP evolution and resummation of
double logarithmic terms $ln^2x$ as well. Using the dynamical GRV input 
parametrization at $k_0^2 = 1 {\rm GeV}^2$ we have found numerically in our 
LO DGLAP + $ln^2x$ approach $q_{val}(x,Q^2)$ at higher scale $Q^2\sim 4m_h^2$ 
and hence the cross section $\sigma (P\bar{P} \rightarrow h\bar{h}) 
- \sigma (PP \rightarrow h\bar{h})(s)$ and  the differential cross section 
$\frac{d}{d\Delta{y}}[\sigma (P\bar{P} \rightarrow h\bar{h}) - 
\sigma (PP \rightarrow h\bar{h})]$ for the charm and bottom production. Very
interesting is an analysis where the rapidity gap $\Delta{y}$ between the quark
and the antiquark is large. The large  rapidity gap ($\Delta{y}>2.0$) causes
that double logarithmic terms $ln^2x\sim \Delta{y}^2$ become important.
Processes with large rapidity gaps in high energy $PP$ and $P\bar{P}$
collisions, observed at the Tevatron \cite{b12} have been theoretically
investigated e.g. in \cite{b13}, \cite{b14}, \cite{b15}, \cite{b16}. The
main reason that the large rapidity gap processes are so intensively studied
is the opportunity to search for new particles e.g. Higgs bosons. Besides,
the hard partonic processes with a rapidity gap between two jets in a final
state enable examination of BFKL Pomeron behaviour \cite{b17}, \cite{b18}.
Even though the pure mesonic exchange process gives a small contribution to
cross sections in high energy hadron - hadron collisions, it could be
interesting because it enable the PQCD test in double logarithmic $ln^2x$
approximation. In our paper we have modified the mesonic Regge picture by
PQCD LO DGLAP method combined with $ln^2x$ terms resummation. We have found
that for the large rapidity gap $\Delta{y}$ between a quark and an antiquark
double logarithmic terms $ln^2x$ become significant. It must be however
emphasized that in an unpolarized case it is very difficult to observe a
dominant role of the $ln^2x$ contribution to the flavour nonsinglet part of
heavy quark production cross sections. There are two reasons of this fact.
First: $ln^2x$ terms in $q_{val}(x,Q^2)$, which by definition are
significant in the small $x$ region ($x\leq 10^{-2}$) are partially hidden
behind leading Regge $x^{\alpha_2(0)}$ behaviour of $q_{val}$. Second: the
quark - antiquark scattering cross section $d\hat\sigma /d\hat{t}$ is
suppressed by a factor of $\hat{s}^{-2}$, so one should minimize $\hat{s}$
with a reasonably large rapidity gap $\Delta{y} = ln(\hat{s}/4m^2)$. Larger
$\Delta{y}$ causes rapid decrease of the suitable cross sections, which
however can be measured at Tevatron and LHC.

\section*{Acknowledgements}

We are grateful to Professor Jan Kwieci\'nski for main idea of this work,
for useful discussions and constructive comments.

\end{document}